\documentclass[12pt]{elsarticle}

\usepackage{titlesec}
\usepackage{enumitem}
\usepackage{hyperref}
\usepackage{amssymb}
\usepackage{algorithm}
\usepackage{algorithmic}

\usepackage{subcaption}

\usepackage[labelfont=bf,labelsep=period]{caption}
\usepackage{multirow}
\usepackage{makecell}

\usepackage{afterpage}
\usepackage{rotating}
\usepackage{subcaption}

\usepackage{lineno}

\usepackage{amsmath}
\usepackage{graphicx}
\usepackage{adjustbox}
\usepackage{xcolor}
\usepackage[utf8]{inputenc} % allow utf-8 input
\usepackage[T1]{fontenc}    % use 8-bit T1 fonts
\usepackage{hyperref}       % hyperlinks
\usepackage{url}            % simple URL typesetting
\usepackage{booktabs}       % professional-quality tables
\usepackage{amsfonts}       % blackboard math symbols
\usepackage{nicefrac}       % compact symbols for 1/2, etc.
\usepackage{microtype}      % microtypography
\usepackage{lipsum}
\usepackage{graphicx}
\graphicspath{{media/}}     % organize your images and other figures under media/ folder
\usepackage{float}
\usepackage{multirow}

\journal{arXiv}

\begin{document}

\begin{frontmatter}


%affiliation{organization={},

\title{Forecasting Commodity Price Shocks Using Temporal and Semantic Fusion of Prices Signals and Agentic Generative AI Extracted Economic News}


\author[label1]{Mohammed-Khalil Ghali} %% Author name
\author[label1]{Cecil Pang} %% Author name
\author[label1]{Oscar Molina}
\author[label1]{Carlos Gershenson-Garcia}
\author[label1]{Daehan Won\corref{cor1}} %% Author name with corresponding author note
\cortext[cor1]{Corresponding author}
\ead{dhwon@binghamton.edu}

%% Author name

\affiliation[label1]{organization={School of Systems Science and Industrial Engineering, State University of New York at Binghamton},%Department and Organization
            addressline={4400 Vestal Pkwy}, 
            city={Binghamton},
            postcode={13902}, 
            state={NY},
            country={USA}}

%% Abstract
\begin{abstract}
Accurate forecasting of commodity price spikes is vital for countries with limited economic buffers, where sudden increases can strain national budgets, disrupt import-reliant sectors, and undermine food and energy security. This paper introduces a hybrid forecasting framework that combines historical commodity price data with semantic signals derived from global economic news, using an agentic generative AI pipeline. The architecture integrates dual-stream Long Short-Term Memory (LSTM) networks with attention mechanisms to fuse structured time-series inputs with semantically embedded, fact-checked news summaries collected from 1960 to 2023.

The model is evaluated on a 64-year dataset comprising normalized commodity price series and temporally aligned news embeddings. Results show that the proposed approach achieves a mean AUC of 0.94 and an overall accuracy of 0.91 substantially outperforming traditional baselines such as logistic regression (AUC = 0.34), random forest (AUC = 0.57), and support vector machines (AUC = 0.47). Additional ablation studies reveal that the removal of attention or dimensionality reduction leads to moderate declines in performance, while eliminating the news component causes a steep drop in AUC to 0.46, underscoring the critical value of incorporating real-world context through unstructured text. These findings demonstrate that integrating agentic generative AI with deep learning can meaningfully improve early detection of commodity price shocks, offering a practical tool for economic planning and risk mitigation in volatile market environments while saving the very high costs of operating a full generative AI agents pipeline.
\end{abstract}

%    \centering
%    \begin{figure}
%        \centering
%        \begin{subfigure}{\textwidth}
%            \centering
%            \includegraphics[width=\linewidth]{images/Methodology (2).png}
%            \caption{GTR model}
%            \label{fig:imageA}
%        \end{subfigure}
%        \vspace{0.5cm} % Space between the images
%        \begin{subfigure}{\textwidth}
%            \centering
%            \includegraphics[width=\linewidth]{images/texttosql.png}
%            \caption{GTR-T model}
%        \end{subfigure}
%        \caption{GTR and GTR-T models graphical abstracts}
%        \label{fig:general}
%    \end{figure}

\begin{highlights}

\item This paper presents a novel hybrid framework combining agentic generative AI with deep learning

\item This paper introduces a dual-stream LSTM model with attention to fuse price data and news embeddings

\item This paper demonstrates strong evaluation with 0.94 mean AUC  and 0.91 accuracy on 64 years of data

\item This paper shows ablation studies confirming semantic news context is critical for accurate forecasts 

\end{highlights}

%% Keywords
\begin{keyword}
Agents \sep Deep Learning \sep Generative AI \sep Commodities Finance \sep World Bank




\end{keyword}

\end{frontmatter}


\section{Introduction}
\label{sec:intro}
Commodity price volatility poses significant challenges to global economic stability, with particularly severe consequences for developing nations. Sudden surges in the prices of essential commodities such as food, energy, and raw materials can disrupt global and local supply chains, strain public finances, and intensify existing issues of poverty, food insecurity, and social unrest \cite{wheeler2020adding}. Developing countries, often characterized by fragile economic structures and a high dependency on imports, are disproportionately affected due to their limited fiscal space and weaker institutional capacities  \cite{de2010agricultural} to manage exogenous shocks.

These price fluctuations arise from a complex interplay of factors, including geopolitical conflicts, climatic disruptions, financial speculation, and evolving demand-supply dynamics in international markets \cite{abbott2011s}. For instance, the COVID-19 pandemic and the Russia-Ukraine conflict have demonstrated how sudden global events can lead to dramatic commodity price surges, significantly affecting import-reliant nations \cite{mamun2024role}. In this environment, accurate and timely forecasting becomes an indispensable component of policy and economic planning. Yet, traditional forecasting models, which predominantly rely on historical time-series data  \cite{ghoshray2014breaks} and econometric approaches, often prove inadequate when faced with nonlinear disruptions and unexpected events.

Emerging technologies, particularly in natural language processing (NLP) and machine learning (ML), offer new possibilities to address these limitations. By enabling the integration of unstructured textual data—such as global news, policy reports, and social media feeds—these technologies can capture real-time market sentiment, early warning signals, and thematic trends that are not reflected in historical prices alone \cite{zhang2022exploring}. Deep learning models, especially those utilizing transformer-based architectures and sentiment-aware networks, have shown promise in financial and commodity markets \cite{kaur2023trade} for enhancing predictive accuracy and adaptability to new information.

For low-income countries, where the margin for error is minimal, adopting intelligent, hybrid decision support systems that fuse numerical indicators with qualitative analysis can provide a strategic edge. Such systems can help anticipate risks, support more informed budget allocations, and enable proactive interventions to shield vulnerable populations from the ripple effects of global price shocks. In particular, integrating generative AI agents that autonomously extract, interpret, and synthesize semantic signals from vast textual corpora may represent a significant step forward in decision making infrastructure.

In this context, this research proposes the development of a hybrid forecasting framework that combines deep learning techniques with semantic analysis of news data using an agentic generative AI-based approach. The system aims to enhance the accuracy and responsiveness of commodity price forecasts, with a specific focus on empowering policymakers in developing economies. By bridging data-driven analytics with contextual intelligence, the proposed framework aspires to be a valuable tool in navigating the uncertainties of global commodity markets.

The remainder of this paper is organized as follows. Section~\ref{sec:literature} reviews existing approaches to commodity price forecasting, highlighting both traditional econometric models and recent developments in deep learning and generative AI. Section~\ref{sec:dataset} describes the dataset construction, including preprocessing of historical price data and semantic news extraction. Section~\ref{sec:methodology} outlines the proposed multimodal forecasting framework that fuses temporal and semantic signals using agentic generative AI. Section~\ref{sec:results} presents the evaluation results and ablation studies. Finally, Section~\ref{sec:conclusion} concludes the paper with key insights, limitations, and future research directions.

\section{Literature Review}
\label{sec:literature}
Forecasting financial and commodity prices has evolved through decades of modeling, beginning with classical time-series techniques and advancing to deep learning and generative AI-driven methods. Traditional models like ARIMA have historically served as strong baselines in short-term commodity price forecasting due to their simplicity and solid statistical foundations \cite{Behmiri2013}. ARIMA models, with their autoregressive and moving average components, are particularly effective when applied to stationary and linear time series data, making them ideal for controlled or stable markets. However, their inherent assumption of linearity and constant variance limits their flexibility, especially in dynamic and non-stationary environments where market shocks, geopolitical factors, and speculative behavior play a significant role. These limitations become even more pronounced in commodity markets, which are characterized by high volatility, seasonality, and abrupt structural changes. Consequently, while ARIMA remains useful as a benchmark model, its linear structure often falls short when compared to non-linear methods such as artificial neural networks \cite{Lasheras2015}, which are better equipped to model complex, nonlinear dependencies. Moreover, Vector Autoregression (VAR) and its Bayesian variants (BVAR) have expanded modeling capacity by allowing for the joint modeling of multiple time series, thereby enabling the inclusion of macroeconomic variables like exchange rates, inventory levels, or stock indices \cite{Kwas2021}. These multivariate approaches offer a richer contextual understanding of commodity dynamics and allow the model to exploit interdependencies among variables. Structural models that explicitly consider fundamental factors such as OPEC decisions, global GDP growth, and supply chain disruptions add a layer of interpretability and policy relevance to the forecasts, making them valuable for decision makers. To further enhance stability and generalization, ensemble approaches that combine forecasts from multiple model classes, such as machine learning, econometrics, and structural models, have shown to yield more robust and accurate predictions \cite{Baumeister2015}.

In parallel, deep learning models, particularly Long Short-Term Memory (LSTM) networks and Gated Recurrent Units (GRU), have demonstrated their superiority in capturing long-range dependencies and complex temporal dynamics in commodity price series \cite{Sezer2020,Zhao2017}. These models are capable of learning intricate, nonlinear relationships without the need for manual feature engineering, making them especially suitable for high-dimensional and noisy financial data. LSTM networks, with their memory cells and gating mechanisms, address the vanishing gradient problem and effectively capture sequential dependencies across different time scales, from short-term fluctuations to long-term trends. GRUs provide a more computationally efficient alternative with similar performance characteristics. Enhancements to these architectures such as hybrid decomposition techniques like Empirical Mode Decomposition (EMD), wavelet transforms, and Discrete Fourier Transforms help in isolating noise and improving signal clarity, which in turn boosts forecasting accuracy across diverse financial datasets \cite{Chiroma2015,Lasheras2015}. The integration of convolutional layers further enriches the feature extraction process by allowing the model to detect local patterns and abrupt shifts in price series. Attention mechanisms, which allow the model to weigh the importance of different input time steps, and Transformer-based models such as the Temporal Fusion Transformer (TFT), have revolutionized sequence modeling. These attention-based architectures selectively attend to the most informative inputs, allowing the model to prioritize relevant historical data points and discard noise, thereby improving both interpretability and forecasting performance \cite{Chen2024}. This flexibility makes them highly effective for modeling complex, nonlinear, and often chaotic market behaviors.

A critical advancement in the field is the integration of unstructured news data into prediction pipelines. While price time series alone may reflect past performance, textual data from news articles, financial headlines, and analyst reports provides real-time insight into investor sentiment, macroeconomic trends, and major geopolitical developments. Empirical studies indicate that media coverage, sentiment polarity, and event-driven signals can significantly enhance model accuracy, especially for volatility forecasting and directional price changes. Tetlock’s seminal work \cite{Tetlock2007} demonstrated that pessimism in financial news could negatively affect equity prices, establishing a foundational link between textual sentiment and market behavior. Building on this, more recent research has shown that incorporating sentiment scores—derived from linguistic analysis of headlines and articles—can improve deep learning-based forecasts for oil and commodity futures \cite{Li2021,Banerjee2024}. These studies reveal that investor psychology and news-driven reactions play a critical role in short-term market dynamics \cite{BOLLEN20111}. Event-based models, which utilize natural language processing (NLP) to identify textual markers of major geopolitical, financial, or supply-chain disruptions, further refine predictions by incorporating context that raw numerical data may miss \cite{Ding2015}. Systematic reviews and meta-analyses confirm that fusing text with price data consistently improves directional forecasting accuracy across various commodities and financial instruments \cite{Nassirtoussi2014,Chen2024}. This multimodal approach enables models to react more adaptively to real-world developments.

With the rise of transformer models, financial NLP has significantly matured. These architectures, originally developed for machine translation and sequence modeling, have been adapted to finance-specific tasks with notable success. FinBERT, a variant of BERT fine-tuned on financial texts, offers state-of-the-art sentiment classification tailored for corporate filings, earnings reports, and analyst notes \cite{Liu2020}. Its domain-specific vocabulary and contextual embeddings allow it to outperform generic models in identifying sentiment nuances unique to financial language. Hybrid models that combine FinBERT-derived embeddings with numerical inputs, such as LSTM-based sequences of past prices, have been shown to outperform models using either modality alone \cite{Chen2024}. These multimodal pipelines leverage the complementary strengths of text and numerical data: textual features capture qualitative shifts in market perception, while price data reflects historical patterns and volatility regimes. On the generative side, large language models such as GPT-3 and GPT-4 have demonstrated impressive capabilities in summarizing financial reports, generating analyst-style commentary, and inferring the directional impact of news and events \cite{LopezLira2023}. These models exhibit few-shot and zero-shot learning capabilities, allowing them to generalize across different market scenarios with minimal fine-tuning. BloombergGPT, a recent domain-specific generative model trained on massive proprietary financial datasets, extends these capabilities to high-frequency and real-time forecasting tasks \cite{Wu2023}. Its architecture and training corpus enable it to handle nuanced financial language, complex reasoning, and real-time event summarization, showcasing the growing feasibility of automated, real-time market analysis using large language models (LLMs).

Generative AI agents \cite{mghali} represent the convergence of LLMs, tool-use, and real-time market analysis. These agents operate as autonomous systems that continuously retrieve, summarize, and interpret evolving market information in a feedback-driven loop. FinAgent, for instance, employs a multi-module architecture that integrates LLMs for reasoning and summarization, memory buffers for context retention, and a market intelligence engine for real-time data access and processing \cite{Zhang2024}. This modular design allows FinAgent to adapt to rapidly changing market environments and make autonomous trading decisions based on the latest information. Similarly, models like TradingGPT and LLMFactor exemplify how generative agents can specialize in narrowly defined tasks, such as sentiment tracking, volatility prediction, or identifying macroeconomic factors influencing commodity prices \cite{Li2023,Wang2024}. These agents are capable of initiating sub-tasks, evaluating evidence, and synthesizing insights to produce actionable intelligence. Retrieval-augmented agents take this capability further by grounding generated outputs in up-to-date, factual information retrieved from trusted financial databases or live news feeds. This grounding significantly enhances robustness, reduces hallucination, and ensures decision making is aligned with current market conditions.

The fusion of textual and numerical data has led to substantial gains in predicting rare but impactful events such as price spikes. Studies show that integrating sentiment indicators or event embeddings with time-series models improves the precision and recall of spike detection \cite{Ding2015,Li2021,Chen2024}. This multimodal approach also benefits volatility forecasting, particularly when traditional models are extended with sentiment-aware conditional variance structures, as in GARCH models. Regime shifts, often triggered by exogenous news events, are better captured when the model accounts for real-time textual cues \cite{Ramyar2019}.

Nonetheless, challenges persist in aligning news with time-series data. Text arrives asynchronously and varies in granularity, making temporal matching and noise filtration non-trivial. Different fusion strategies like early (input-level), intermediate (representation-level), and late (decision level) carry trade-offs in complexity and interpretability \cite{Jiang2024}. Moreover, the black box nature of deep models complicates transparency. While attention maps and explainability techniques like SHAP offer partial solutions, end-users often require more interpretable rationales, especially in financial domains. Lastly, real-time constraints impose latency and scalability limitations on LLM-powered systems, reinforcing the need for efficient and robust design pipelines.
\newline
In response to these opportunities and challenges, the proposed framework introduces:
\begin{itemize}
\item A hybrid deep learning model that combines LSTM-based price forecasting with attention mechanisms is developed to integrate contextual information from financial news with commodities prices data.
\item A generative AI agentic approach is introduced to autonomously retrieve, filter, and summarize relevant news from major global news outlets in real-time, providing targeted and timely inputs to the forecasting pipeline.
\item The model is evaluated on a comprehensive dataset comprising 64 years of global commodity prices and aligned news records, using multiple rigorous validation metrics to assess robustness and generalization.
\end{itemize}

\begin{table}[H]
\centering
\caption{Summary of related work and comparative contributions}
\renewcommand{\arraystretch}{1.4}
\begin{tabular}{>{\raggedright\arraybackslash}p{3.5cm} p{5.5cm} p{5.5cm}}
\hline
\textbf{Methodology} & \textbf{Approach and Strengths} & \textbf{Limitations and Proposed Model Improvements} \\
\hline
ARIMA, VAR, BVAR \cite{Behmiri2013,Kwas2021,Baumeister2015} & Simple and interpretable; well-suited for short-term and low-frequency data analysis. & Limited in capturing non-linearity or incorporating exogenous signals. The proposed model handles both nonlinear and multimodal inputs. \\
\hline
LSTM, GRU-based Deep Models \cite{Sezer2020,Zhao2017,Chiroma2015} & Effectively captures temporal dependencies and patterns in dynamic time-series data. & Lacks awareness of external context like news events. The proposed model integrates news-based context for enhanced accuracy. \\
\hline
Sentiment-Aware Models (e.g., FinBERT+LSTM) \cite{Liu2020,Chen2024} & Leverages semantic insights from financial text to improve forecast quality. & Typically limited to static embeddings or pre-aggregated news. A real-time generative news retrieval and fusion is introduced. \\
\hline
Transformer-based Models \cite{Chen2024} & Models long-range dependencies and applies attention for improved interpretability. & High computational cost and less flexibility with unstructured external data. The proposed hybrid design handles structured and unstructured fusion efficiently. \\
\hline
Generative AI Agents (e.g., TradingGPT, Fin-Agent) \cite{Zhang2024,Li2023,Wang2024} & Automatically retrieve, summarize, and synthesize real-time financial insights. & Largely focused on equity markets. The proposed model targets global commodities and incorporates long-term historical context. \\
\hline
\textbf{The proposed model} & Real-time hybrid forecasting that fuses LSTM, attention, and generative AI agentic news retrieval. & Enables early spike detection with robust generalization across 64 years of commodity and news data. \\
\hline
\end{tabular}
\end{table}

\section{Dataset}
\label{sec:dataset}

This study utilizes publicly available historical commodity price data provided by the World Bank \cite{worldbank2024commodities}, covering the period from 1960 to 2023. The dataset includes yearly average prices for a wide range of commodities, serving as the foundational source for constructing both input features and target labels for the forecasting model. Each year is associated with a computed average price, and these values are used to determine significant variations in commodity trends.

To identify spike events, year-over-year percentage changes are calculated. A year $y_i$ is labeled as a spike year if the relative change in average price from the previous year exceeds a predefined threshold. Specifically, the binary spike label $s_i$ is assigned as:
\begin{equation}
s_i =
\begin{cases}
1 & \text{if } \frac{\text{AvgPrice}_{i} - \text{AvgPrice}_{i-1}}{\text{AvgPrice}_{i-1}} \times 100 > 25\% \\
0 & \text{otherwise}
\end{cases}
\end{equation}
This rule corresponds to a 25\% increase in average price over the prior year and serves as a supervised learning target in the classification model.

\begin{figure}[H]
    \centering
    \includegraphics[width=\textwidth]{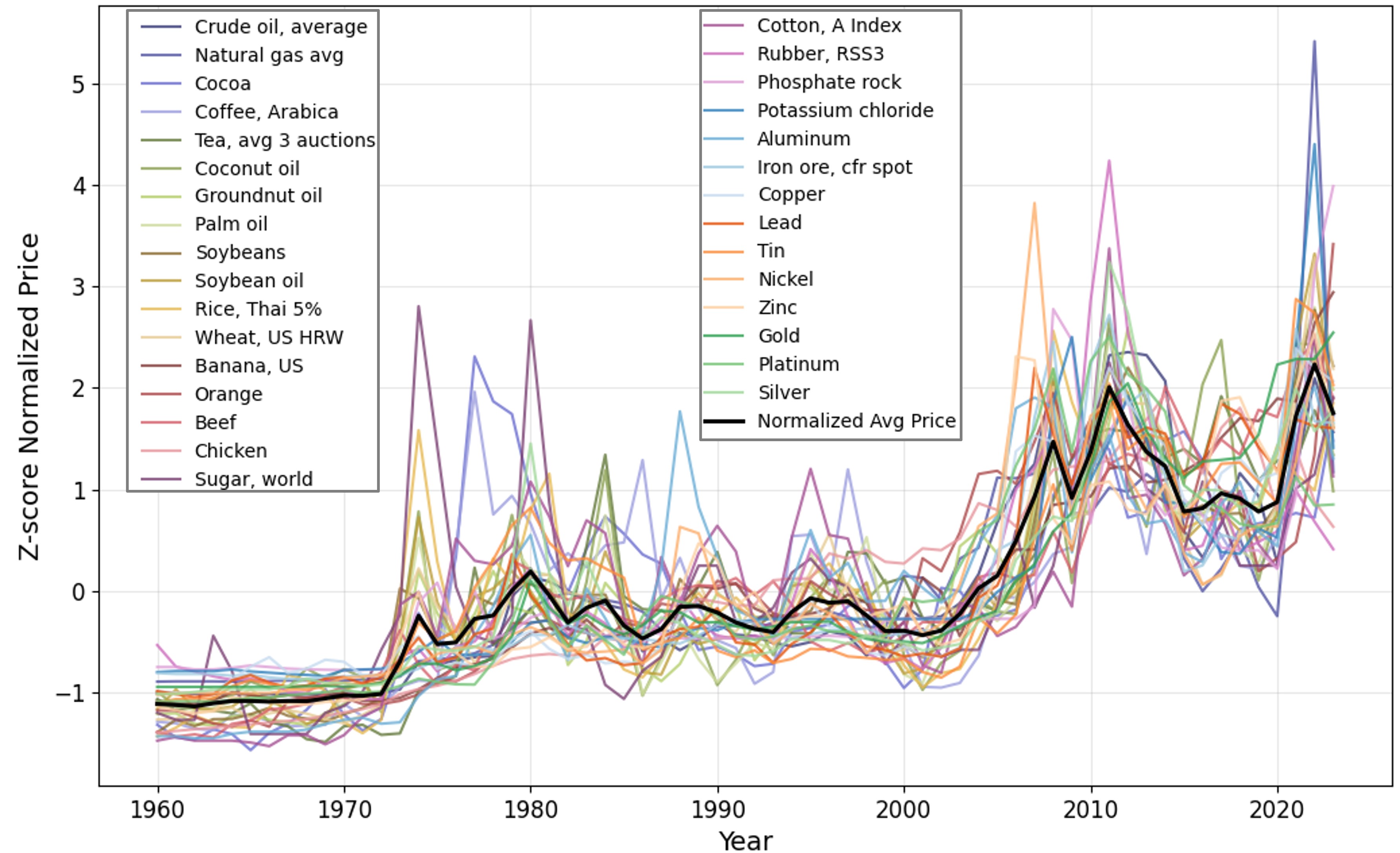}
    \caption{Normalized commodities prices over time with spikes marked}
    \label{fig:allcommodities}
\end{figure}

To ensure comparability across heterogeneous commodity price series that differ in units and magnitude (e.g., \$/bbl for oil versus \$/kg for coffee), z-score normalization was applied to all price columns prior to analysis. This transformation standardizes each commodity’s historical price distribution, enabling a unified analytical scale while preserving relative temporal patterns. Specifically, for each commodity \( i \) and year \( t \), the normalized price \( z_{i,t} \) is computed as:

\begin{equation}
    z_{i,t} = \frac{x_{i,t} - \mu_i}{\sigma_i}
\end{equation}

where \( x_{i,t} \) denotes the raw price, \( \mu_i \) is the historical mean, and \( \sigma_i \) is the standard deviation of commodity \( i \) over the full observation period. This approach allows for the aggregation and direct comparison of commodities in a composite average, which is crucial for detecting broad market trends and identifying spike events based on relative price deviations rather than absolute levels.
Fig. \ref{fig:allcommodities} shows z-score trends for all commodities and detected spikes throughout the dataset period of time. 

\begin{figure}[H]
    \centering
    \includegraphics[width=\textwidth]{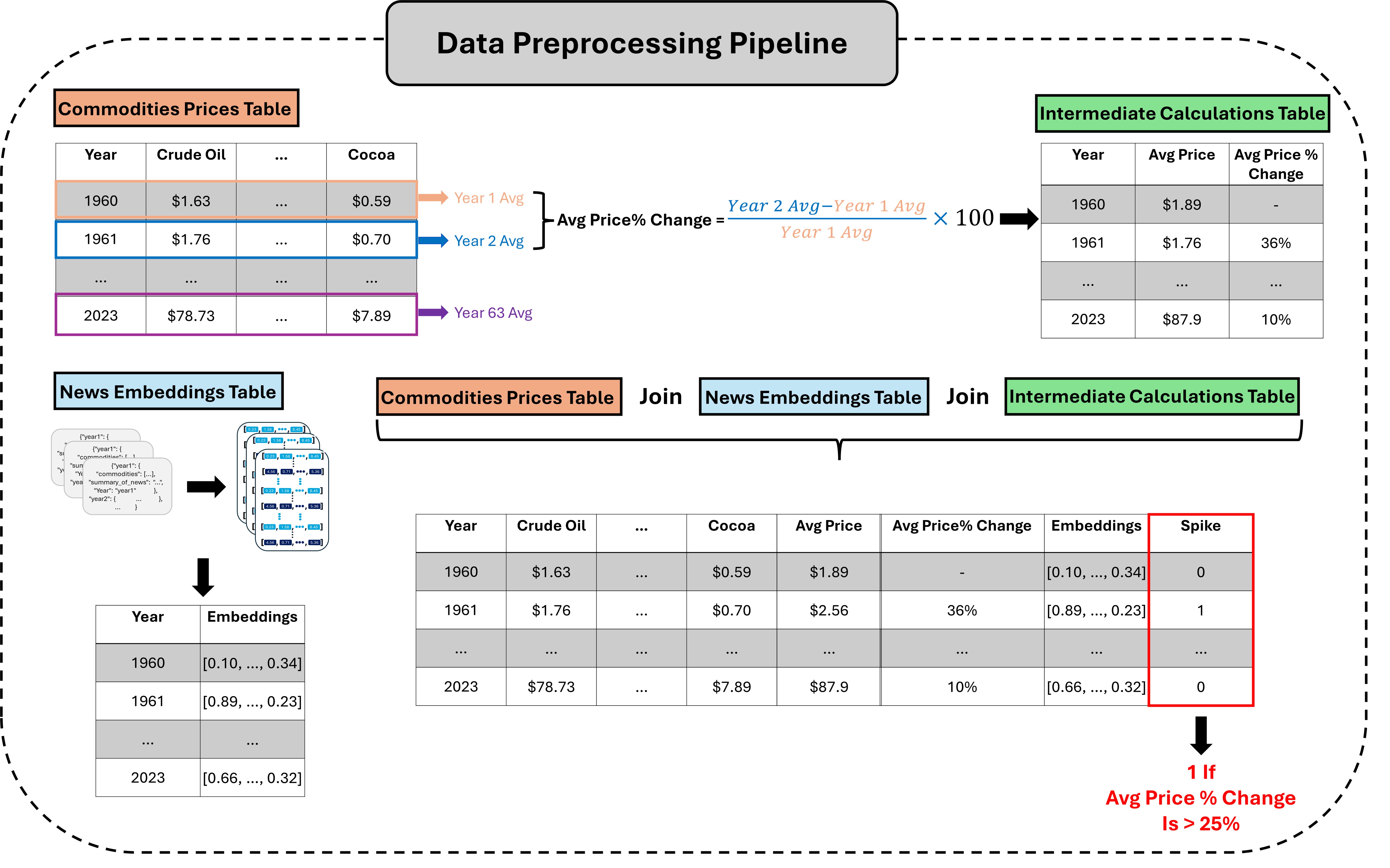}
    \caption{Data preprocessing pipeline with construction of average price time series, spike labels, and embedding-aligned features.}
    \label{fig:preprocessing}
\end{figure}

The overall preprocessing workflow is illustrated in fig. \ref{fig:preprocessing}, which summarizes the data engineering pipeline used to prepare the model-ready dataset. The pipeline begins by extracting yearly average prices from the raw World Bank commodities table. These values are then used to compute yearly percentage changes and assign binary spike labels, which are stored in an intermediate calculations table. Simultaneously, economic news summaries generated through the agentic generative AI pipeline are converted into dense vector representations using sentence-level embedding models. These embeddings are stored in a separate news embeddings table and aligned temporally with their corresponding price and spike labels.

The output of this process is a consolidated dataset consisting of three components: (i) a vector representation of news for each year, (ii) the corresponding average commodity price, and (iii) the binary spike indicator. This structured input is used to train and evaluate the forecasting model described in the methodology section.

\section{Methodology}
\label{sec:methodology}
\begin{figure}[H]
	\centering
	\includegraphics[width=\textwidth]{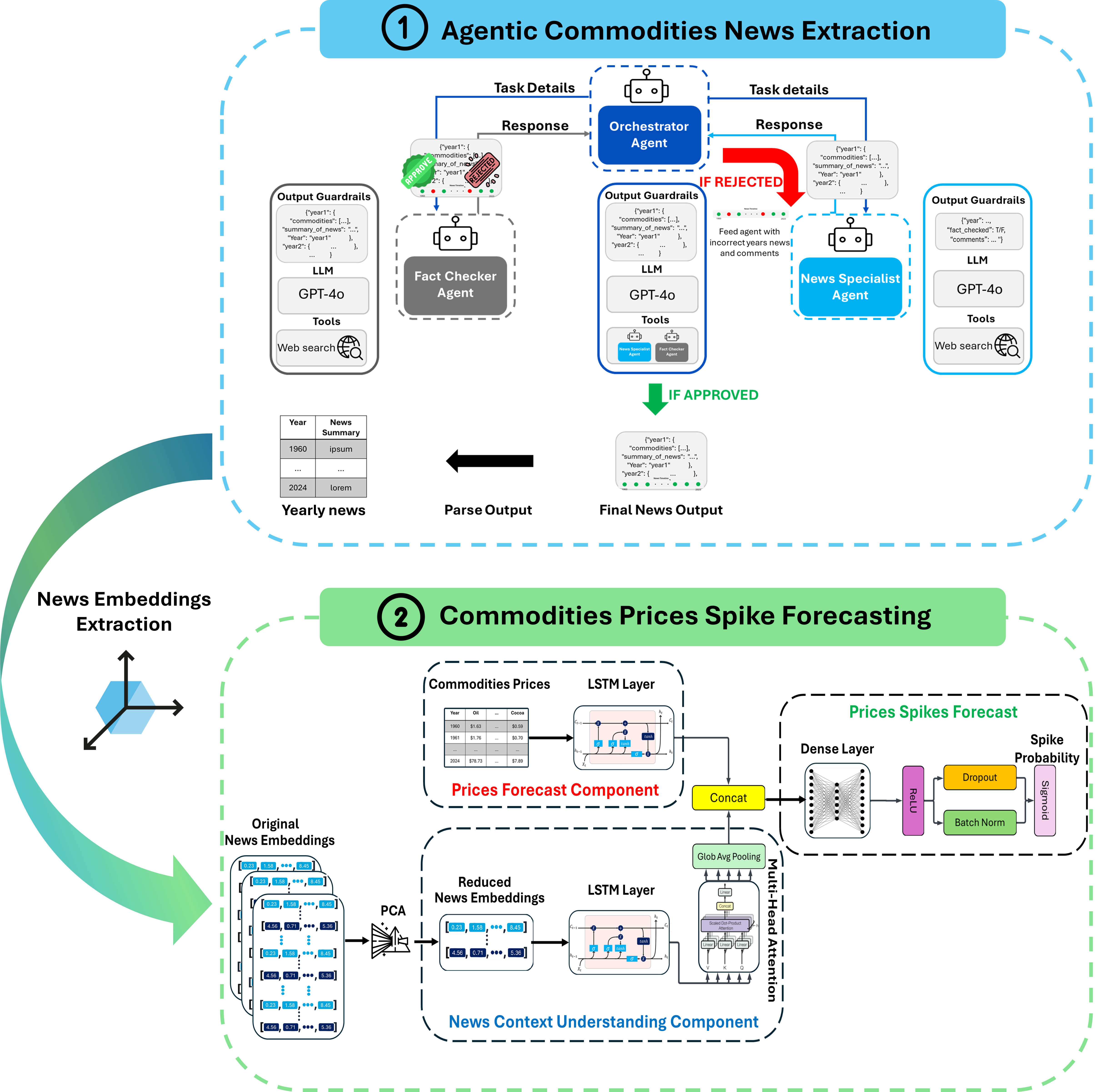}
	\caption{Double phase commodities price spike forecast graphical abstract}\label{graphicalabstract}
\end{figure}

Fig. \ref{graphicalabstract} describes the model proposed to predict price shocks by fusion of agentic generative AI, deep learning price forecast by semantic enrichment using attention mechanisms. The first component extracts commodities related news that will be used by the second component to enrich the context of price shocks forecast. Each of these components are described in details in the following subsections
\subsection{Agentic Generative AI News Extraction}

\begin{figure}[H]
	\centering
	\includegraphics[width=0.85\textwidth]{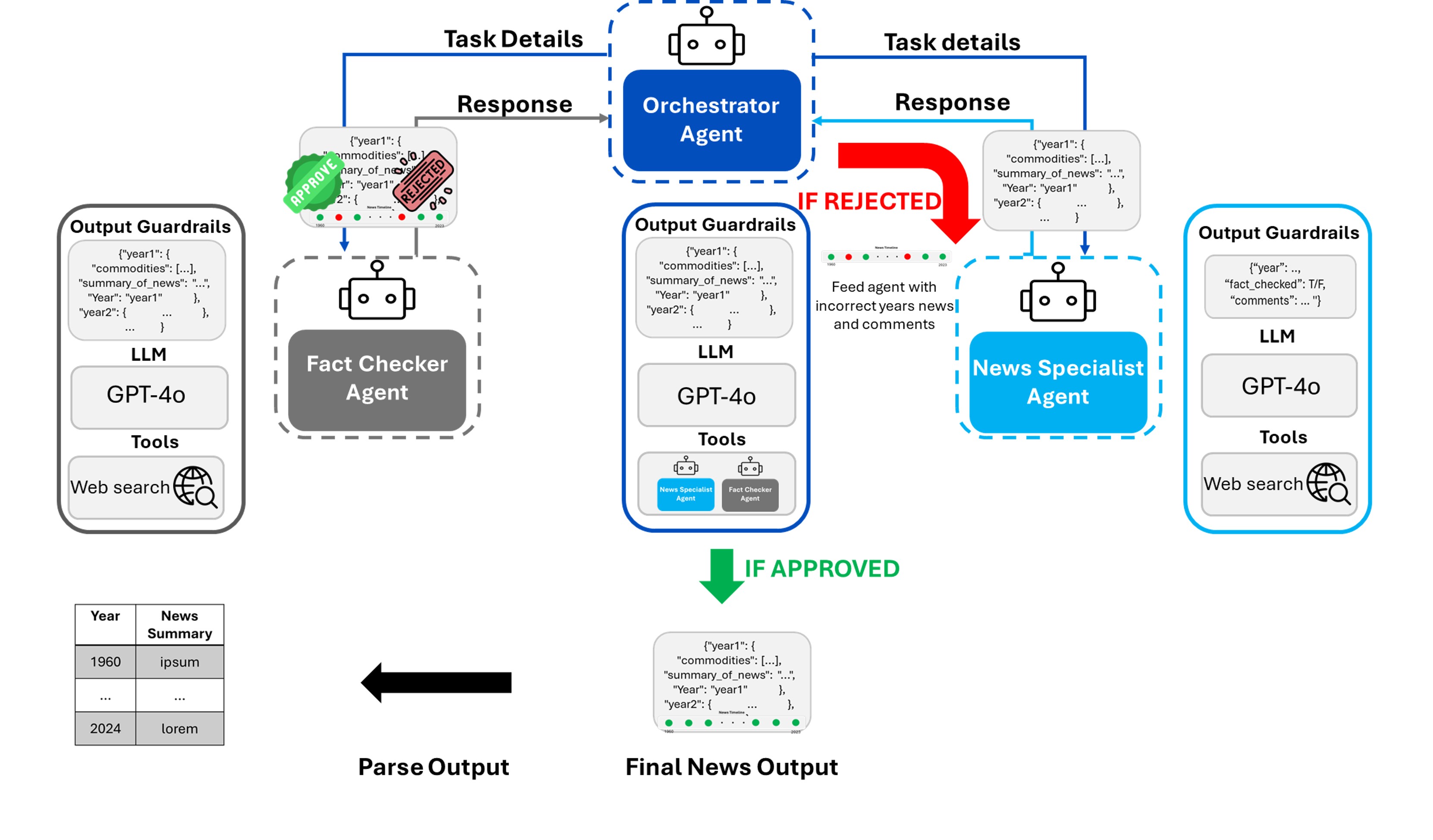}
	\caption{News extraction using an agentic generative AI approach}\label{agenticframework}
\end{figure}

To construct a temporally aligned dataset of economic news summaries relevant to commodity price movements, a hierarchical agent-based generative framework is employed. OpenAI Agents SDK is used to implement the needed agentic framework\cite{Openai}. The architecture follows a manager–worker paradigm, where a high-level orchestration agent governs the interaction between two specialized sub-agents: a summarization agent responsible for extracting yearly commodity-related news, and a fact-checking agent that ensures factual accuracy. This generative pipeline produces verified yearly summaries spanning from 1960 to 2023, which are subsequently transformed into semantic embeddings for use in the forecasting model described in the following section, fig. \ref{agenticframework} describes the proposed methodology.

Let $\mathcal{Y} = \{y_1, y_2, \dots, y_N\}$ denote the set of years under analysis, with $N=64$ in the current study. For each year $y_i \in \mathcal{Y}$, the orchestrator agent $\mathcal{A}^{\text{mgr}}$ initiates the generation of a draft news summary $\hat{s}_i$ by querying the news specialist agent $\mathcal{A}^{\text{spec}}$, formally expressed as:
\begin{equation}
\hat{s}_i = \mathcal{A}^{\text{spec}}(y_i)
\end{equation}
The generated summary $\hat{s}_i$ is then passed to the fact-checking agent $\mathcal{A}^{\text{fact}}$, which returns a binary verdict $v_i \in \{0, 1\}$ indicating whether the content is factually consistent with historical records:
\begin{equation}
v_i = \mathcal{A}^{\text{fact}}(\hat{s}_i)
\end{equation}
If the verdict $v_i$ equals 0, the manager agent prompts a regeneration of $\hat{s}_i$ until a valid version is obtained. Once a summary passes verification ($v_i = 1$), the final version $s_i$ is accepted:
\begin{equation}
s_i = \hat{s}_i \quad \text{if } v_i = 1
\end{equation}
This iterative procedure is applied for all $y_i \in \mathcal{Y}$, resulting in a structured collection of summaries defined by the dictionary:
\begin{equation}
\mathcal{S} = \{ y_i \mapsto s_i \mid i = 1, 2, \dots, N \}
\end{equation}
Each entry $s_i$ contains the year, a list of referenced commodities, and a concise summary of major economic, geopolitical, and market-related developments impacting commodity prices in that year.
The verified summaries $\{s_i\}_{i=1}^N$ serve as the textual input for semantic embedding. Each summary $s_i$ is transformed into a dense vector $\mathbf{e}_i \in \mathbb{R}^d$ using a foundation language model encoder, denoted as:
\begin{equation}
\mathbf{e}_i = f_{\text{embed}}(s_i)
\end{equation}
These embeddings $\{\mathbf{e}_i\}_{i=1}^N$ are later temporally aligned with price records and reduced in dimensionality via PCA before being integrated with the time-series forecasting model, as detailed in the next section.

Alg. \ref{alg:agentic_news_extraction} shows in details the news extraction framework using agentic generative AI.

\begin{algorithm}[H]
\caption{Agentic generative AI news extraction framework}
\label{alg:agentic_news_extraction}
\begin{algorithmic}
\REQUIRE Set of years $\mathcal{Y} = \{y_1, y_2, \ldots, y_N\}$, where $N = 64$
\REQUIRE Manager agent $\mathcal{A}^{\text{mgr}}$, News specialist agent $\mathcal{A}^{\text{spec}}$, Fact-checking agent $\mathcal{A}^{\text{fact}}$
\REQUIRE Maximum retry attempts $M_{\text{max}} = 5$
\ENSURE Verified news summaries dictionary $\mathcal{S} = \{y_i \mapsto s_i \mid i = 1, 2, \ldots, N\}$
\ENSURE Semantic embeddings $\{\mathbf{e}_i\}_{i=1}^N$ where $\mathbf{e}_i \in \mathbb{R}^d$

\STATE \textbf{Initialize:} $\mathcal{S} \leftarrow \emptyset$, $\mathcal{E} \leftarrow \emptyset$
\FOR{each year $y_i \in \mathcal{Y}$}
    \STATE $\text{retry\_count} \leftarrow 0$
    \STATE $\text{verified} \leftarrow \text{False}$
    \WHILE{$\text{verified} = \text{False}$ \AND $\text{retry\_count} < M_{\text{max}}$}
        \STATE \textbf{Step 1:} Generate draft summary
        \STATE $\hat{s}_i \leftarrow \mathcal{A}^{\text{spec}}(y_i)$ \COMMENT{Query news specialist agent}
        
        \STATE \textbf{Step 2:} Fact-check generated summary
        \STATE $v_i \leftarrow \mathcal{A}^{\text{fact}}(\hat{s}_i)$ \COMMENT{Binary verdict: $v_i \in \{0, 1\}$}
        
        \IF{$v_i = 1$}
            \STATE $s_i \leftarrow \hat{s}_i$ \COMMENT{Accept verified summary}
            \STATE $\text{verified} \leftarrow \text{True}$
            \STATE $\mathcal{S}[y_i] \leftarrow s_i$ \COMMENT{Store in dictionary}
        \ELSE
            \STATE $\text{retry\_count} \leftarrow \text{retry\_count} + 1$
        \ENDIF
    \ENDWHILE
    
    \IF{$\text{verified} = \text{False}$}
        \STATE \textbf{Warning:} Maximum retries exceeded for year $y_i$
        \STATE Use fallback summary or skip year
    \ENDIF
    
    \STATE \textbf{Step 3:} Generate semantic embedding
    \IF{$y_i \in \mathcal{S}$}
        \STATE $\mathbf{e}_i \leftarrow f_{\text{embed}}(s_i)$ \COMMENT{Transform to dense vector $\mathbf{e}_i \in \mathbb{R}^d$}
        \STATE $\mathcal{E} \leftarrow \mathcal{E} \cup \{\mathbf{e}_i\}$
    \ENDIF
\ENDFOR
\RETURN $\mathcal{S}$, $\mathcal{E} = \{\mathbf{e}_i\}_{i=1}^N$
\end{algorithmic}
\end{algorithm}

\subsection{Price Spikes Forecasting}
\begin{figure}[H]
	\centering
	\includegraphics[width=0.4\textwidth]{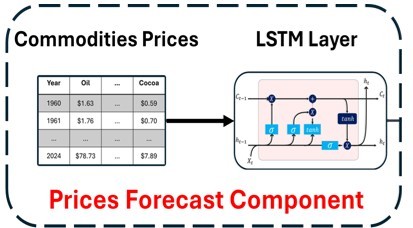}
	\caption{Price forecast component}\label{price_layer}
\end{figure}

\begin{figure}[H]
	\centering
	\includegraphics[width=0.4\textwidth]{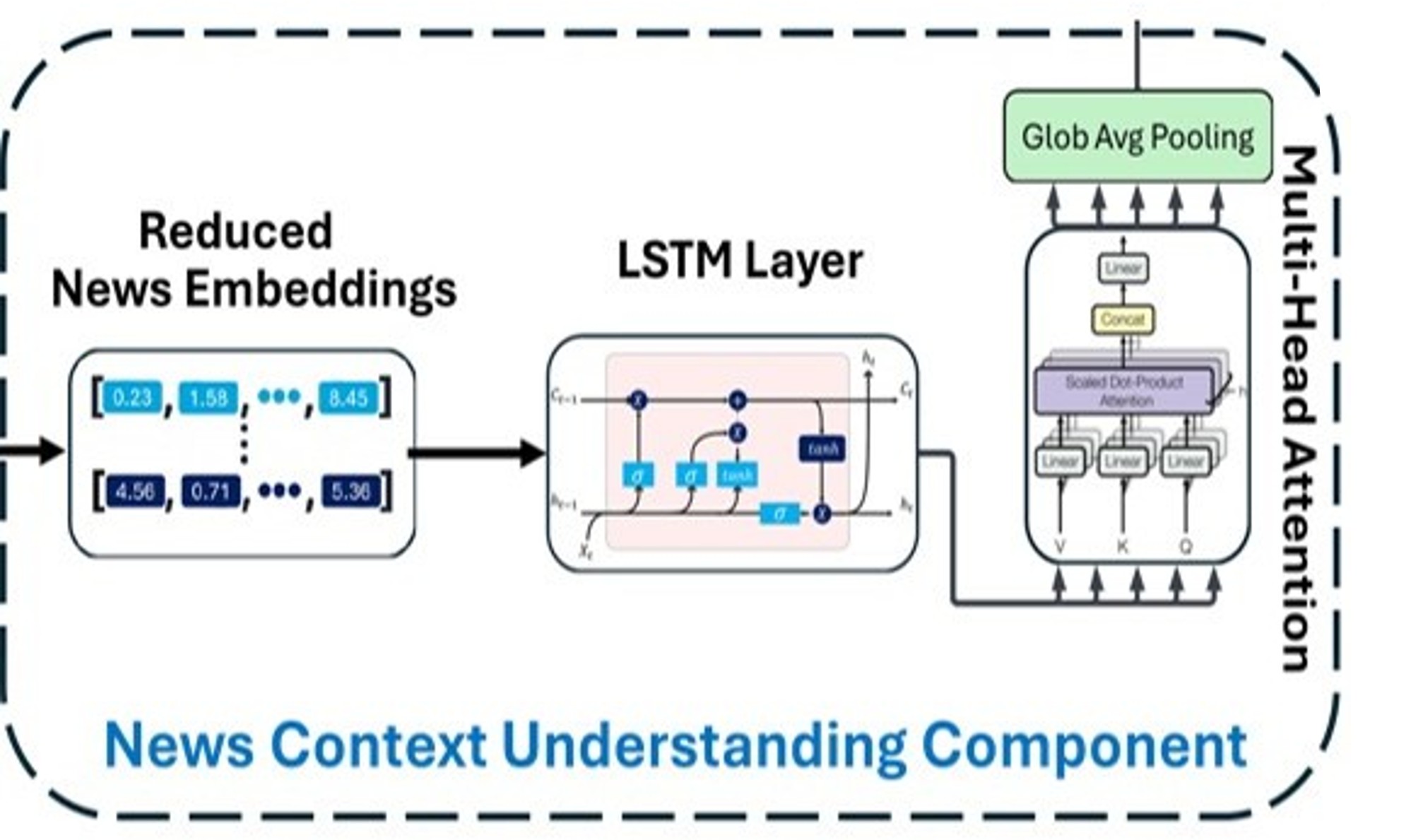}
	\caption{News context component}\label{news_layer}
\end{figure}

\begin{figure}[H]
	\centering
	\includegraphics[width=0.4\textwidth]{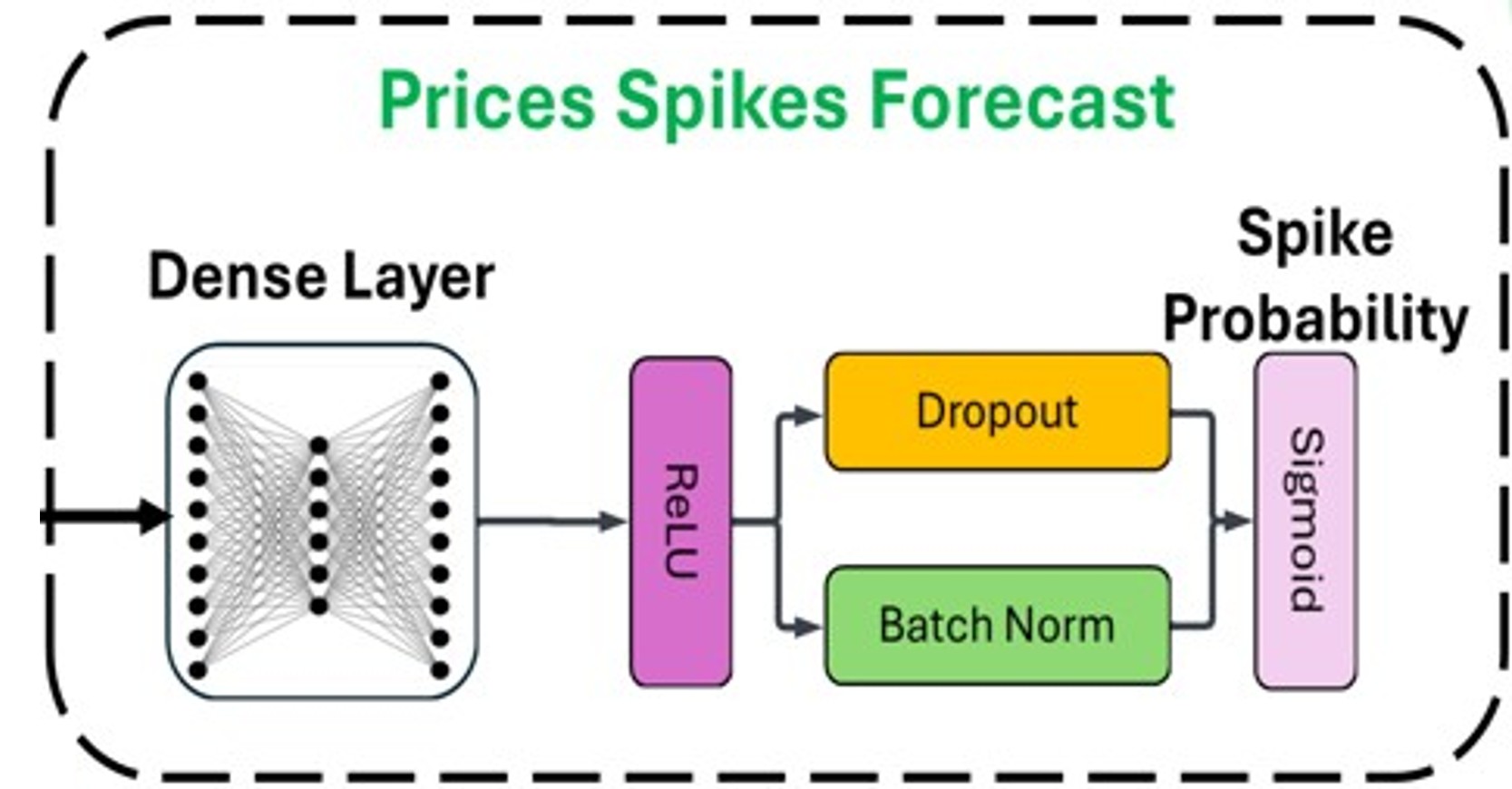}
	\caption{Price spikes forecast}\label{price_forecca}
\end{figure}

To forecast commodity price shocks, a deep learning and transformer-based architecture is proposed that integrates historical price dynamics with semantic representations derived from financial news. The architecture adopts a dual-stream LSTM design with attention mechanisms, enabling parallel processing of structured time-series data and unstructured text embeddings. The objective is to predict the occurrence of a price spike at the next time step based on recent patterns in both modalities.
It starts with a dimensionality reduction of news embeddings where each news headline is first encoded into a high-dimensional vector $\mathbf{e}_t \in \mathbb{R}^d$ using an LLM. To reduce computational overhead and mitigate overfitting risks, Principal Component Analysis (PCA) is applied to obtain a compact representation:
\begin{equation}
\tilde{\mathbf{e}}_t = \mathbf{W}_{\text{PCA}}^\top \mathbf{e}_t, \quad \tilde{\mathbf{e}}_t \in \mathbb{R}^{d'}, \quad d' \ll d
\end{equation}
where $\mathbf{W}_{\text{PCA}}$ contains the top $d'$ principal components extracted from the training set.
To capture sequential dependencies, a fixed-length sliding window is applied to both the price series and the reduced news embeddings. For each time step $t$, the following sequences are defined:
\begin{align}
\mathbf{P}_t &= [p_{t-k+1}, \dots, p_t]^\top \in \mathbb{R}^{k \times 1} \\
\mathbf{E}_t &= [\tilde{\mathbf{e}}_{t-k+1}, \dots, \tilde{\mathbf{e}}_t]^\top \in \mathbb{R}^{k \times d'}
\end{align}
Here, $k$ denotes the window size, and the binary target variable $y_{t+1} \in \{0, 1\}$ indicates whether a price spike occurs at the next time step.
The model consists of two parallel branches. The price sequence $\mathbf{P}_t$ is processed using a unidirectional LSTM as shown in fig. \ref{price_layer} to produce a fixed-length latent representation:
\begin{equation}
\mathbf{h}^{\text{price}}_t = \text{LSTM}_{\text{price}}(\mathbf{P}_t) \in \mathbb{R}^{h}
\end{equation}
\noindent The news sequence $\mathbf{E}_t$ is fed into a second LSTM, followed by a scaled dot-product attention mechanism as shown in fig. \ref{news_layer}, the LSTM output is denoted by:
\begin{equation}
\mathbf{H}^{\text{news}}_t = \text{LSTM}_{\text{news}}(\mathbf{E}_t) \in \mathbb{R}^{k \times h}
\end{equation}
From this output, query, key, and value matrices are computed:
\begin{equation}
\mathbf{Q}_t = \mathbf{H}^{\text{news}}_t \mathbf{W}_Q, \quad
\mathbf{K}_t = \mathbf{H}^{\text{news}}_t \mathbf{W}_K, \quad
\mathbf{V}_t = \mathbf{H}^{\text{news}}_t \mathbf{W}_V
\end{equation}
where $\mathbf{W}_Q, \mathbf{W}_K, \mathbf{W}_V \in \mathbb{R}^{h \times h_a}$ are trainable parameters. Attention weights are computed as:
\begin{equation}
\mathbf{A}_t = \text{softmax} \left( \frac{\mathbf{Q}_t \mathbf{K}_t^\top}{\sqrt{h_a}} \right)
\end{equation}
The attention-based news context vector is defined by:
\begin{equation}
\mathbf{h}^{\text{news}}_t = \frac{1}{k} \sum_{i=1}^{k} \sum_{j=1}^{k} A_{ij} \cdot \mathbf{V}_{t,j}
\end{equation}
The latent representations from the two branches are concatenated and passed through a fully connected layer as shown in fig. \ref{price_forecca}:
\begin{align}
\mathbf{h}^{\text{fused}}_t &= [\mathbf{h}^{\text{price}}_t; \mathbf{h}^{\text{news}}_t] \\
\mathbf{z}_t &= \text{Dropout} \left( \text{ReLU}(\mathbf{W}_1 \mathbf{h}^{\text{fused}}_t + \mathbf{b}_1) \right) \\
\hat{y}_{t+1} &= \sigma(\mathbf{w}_2^\top \mathbf{z}_t + b_2)
\end{align}
\noindent where $\sigma(\cdot)$ is the sigmoid activation function, and $\hat{y}_{t+1}$ represents the probability of a price spike.
The model is trained using the binary cross-entropy loss:
\begin{equation}
\mathcal{L} = -\frac{1}{N} \sum_{t=1}^{N} \left[ y_{t+1} \log \hat{y}_{t+1} + (1 - y_{t+1}) \log (1 - \hat{y}_{t+1}) \right]
\end{equation}
Optimization is performed using the Adam optimizer while $\ell_2$ regularization is applied to the dense layers to improve generalization and prevent overfitting. Alg. \ref{alg:price_shock_forecasting} presents the process of forecasting price shocks in details using deep learning and attention mechanism. 
\newpage

\vspace*{-3cm}
\begin{algorithm}[H]
\caption{Deep learning price shocks forecasting with Semantic Enrichment}
\label{alg:price_shock_forecasting}
\resizebox{\textwidth}{!}{ % Scale to fit page width
\begin{minipage}{\textwidth}
\footnotesize
\begin{algorithmic}
\REQUIRE Historical price series $\{p_t\}_{t=1}^T$, News embeddings $\{\mathbf{e}_t\}_{t=1}^T$ where $\mathbf{e}_t \in \mathbb{R}^d$
\REQUIRE Window size $k$, Reduced embedding dimension $d' \ll d$
\REQUIRE Binary spike labels $\{y_t\}_{t=k+1}^T$ where $y_t \in \{0, 1\}$
\REQUIRE Learning rate $\alpha$, Batch size $B$, Number of epochs $E$
\ENSURE Trained model parameters $\Theta = \{\mathbf{W}_{\text{PCA}}, \theta_{\text{LSTM}}, \mathbf{W}_Q, \mathbf{W}_K, \mathbf{W}_V, \mathbf{W}_1, \mathbf{w}_2, \mathbf{b}_1, b_2\}$
\ENSURE Price shock forecasts $\{\hat{y}_{t+1}\}_{t=k}^{T-1}$

\STATE \textbf{Phase I: Preparation}
\STATE \textbf{Step 1:} Dimensionality reduction of news embeddings
\STATE Compute PCA transformation: $\mathbf{W}_{\text{PCA}} \leftarrow \text{PCA}(\{\mathbf{e}_t\}_{t=1}^T, d')$
\FOR{$t = 1$ to $T$}
    \STATE $\tilde{\mathbf{e}}_t \leftarrow \mathbf{W}_{\text{PCA}}^\top \mathbf{e}_t$ \COMMENT{$\tilde{\mathbf{e}}_t \in \mathbb{R}^{d'}$}
\ENDFOR

\STATE \textbf{Step 2:} Create sliding window sequences
\FOR{$t = k$ to $T-1$}
    \STATE $\mathbf{P}_t \leftarrow [p_{t-k+1}, p_{t-k+2}, \ldots, p_t]^\top$ \COMMENT{Price sequence}
    \STATE $\mathbf{E}_t \leftarrow [\tilde{\mathbf{e}}_{t-k+1}, \tilde{\mathbf{e}}_{t-k+2}, \ldots, \tilde{\mathbf{e}}_t]^\top$ \COMMENT{News sequence}
\ENDFOR

\STATE \textbf{Phase II: Deep Neural Network Training}
\STATE \textbf{Initialize:} All model parameters $\Theta$ randomly
\FOR{epoch $e = 1$ to $E$}
    \STATE Shuffle training data: $\{(\mathbf{P}_t, \mathbf{E}_t, y_{t+1})\}_{t=k}^{T-1}$
    \FOR{each mini-batch of size $B$}
        \STATE \textbf{Forward Pass:}
        \FOR{each sample $(\mathbf{P}_t, \mathbf{E}_t, y_{t+1})$ in batch}
            \STATE \textbf{Step 3:} Process price sequence
            \STATE $\mathbf{h}^{\text{price}}_t \leftarrow \text{LSTM}_{\text{price}}(\mathbf{P}_t; \theta_{\text{LSTM}})$
            
            \STATE \textbf{Step 4:} Process news sequence with attention
            \STATE $\mathbf{H}^{\text{news}}_t \leftarrow \text{LSTM}_{\text{news}}(\mathbf{E}_t; \theta_{\text{LSTM}})$ \COMMENT{$\mathbf{H}^{\text{news}}_t \in \mathbb{R}^{k \times h}$}
            
            \STATE \textbf{Step 5:} Compute attention
            \STATE $\mathbf{Q}_t \leftarrow \mathbf{H}^{\text{news}}_t \mathbf{W}_Q$, $\mathbf{K}_t \leftarrow \mathbf{H}^{\text{news}}_t \mathbf{W}_K$, $\mathbf{V}_t \leftarrow \mathbf{H}^{\text{news}}_t \mathbf{W}_V$
            \STATE $\mathbf{A}_t \leftarrow \text{softmax}\left(\frac{\mathbf{Q}_t \mathbf{K}_t^\top}{\sqrt{h_a}}\right)$ \COMMENT{Attention weights}
            \STATE $\mathbf{h}^{\text{news}}_t \leftarrow \frac{1}{k} \sum_{i=1}^{k} \sum_{j=1}^{k} A_{ij} \cdot \mathbf{V}_{t,j}$ \COMMENT{Weighted context vector}
            
            \STATE \textbf{Step 6:} Fusion and forecast
            \STATE $\mathbf{h}^{\text{fused}}_t \leftarrow [\mathbf{h}^{\text{price}}_t; \mathbf{h}^{\text{news}}_t]$ \COMMENT{Concatenate representations}
            \STATE $\mathbf{z}_t \leftarrow \text{Dropout}(\text{ReLU}(\mathbf{W}_1 \mathbf{h}^{\text{fused}}_t + \mathbf{b}_1))$
            \STATE $\hat{y}_{t+1} \leftarrow \sigma(\mathbf{w}_2^\top \mathbf{z}_t + b_2)$ \COMMENT{Sigmoid activation}
        \ENDFOR
        
        \STATE \textbf{Step 7:} Compute loss and backpropagate
        \STATE $\mathcal{L}_{\text{batch}} \leftarrow -\frac{1}{B} \sum_{i=1}^{B} [y_i \log \hat{y}_i + (1-y_i) \log(1-\hat{y}_i)]$
        \STATE $\nabla_\Theta \mathcal{L}_{\text{batch}} \leftarrow \text{Backpropagation}(\mathcal{L}_{\text{batch}})$
        \STATE $\Theta \leftarrow \Theta - \alpha \nabla_\Theta \mathcal{L}_{\text{batch}}$ \COMMENT{Parameter update}
    \ENDFOR
    
    \STATE \textbf{Step 8:} Validation and early stopping
    \STATE Compute validation loss $\mathcal{L}_{\text{val}}$ on held-out data
    \IF{$\mathcal{L}_{\text{val}}$ stops improving for patience epochs}
        \STATE \textbf{break} \COMMENT{Early stopping}
    \ENDIF
\ENDFOR
\RETURN Probability of price shock at time $t+1$: $\{\hat{y}_{t+1}\}_{t=k}^{T-1}$
\end{algorithmic}
\end{minipage}
}
\end{algorithm}
\vspace{-1em}

\section{Results and Discussion}
\label{sec:results}

This section presents the evaluation outcomes of the proposed multimodal forecasting framework. 
To ensure a rigorous and interpretable evaluation of the proposed multimodal forecasting framework, a two-type validation strategy tailored for temporal data is employed. First, a strict 20\% hold-out test set, never seen by the model during training or validation, is reserved to assess the model’s robustness and generalization capabilities. This final test set is used exclusively for evaluating the model’s ability to anticipate real-world commodity price spikes.
In parallel, a time-aware cross-validation scheme using TimeSeriesSplit is used to perform ablation studies and comparative assessments with baseline models, including Logistic Regression, Support Vector Machines (SVM), and Random Forests. This approach respects the temporal structure of the data and avoids information leakage, enabling reliable performance comparisons under conditions that reflect realistic forecasting constraints. Together, these strategies aim to offer a comprehensive view of the model's performance and the importance of each component.

\begin{figure}[H]
    \centering
    \includegraphics[width=0.45\textwidth]{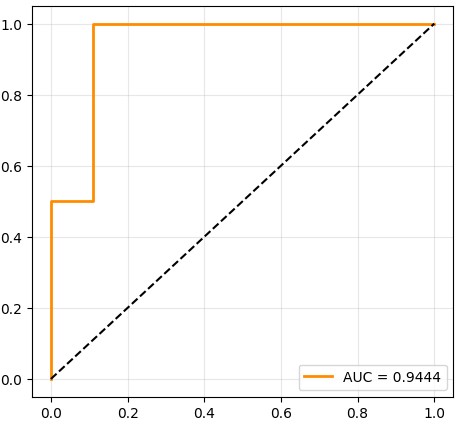}
    \hfill
    \includegraphics[width=0.43\textwidth]{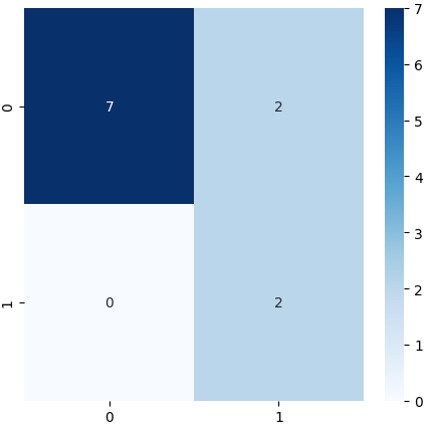}
    \caption{Performance of the full model. Left: ROC curve showing strong classification ability (AUC = 0.9444). Right: Confusion matrix reflecting low false positives and false negatives on testing set.}
    \label{fig:model_performance}
\end{figure}

The ROC curve in fig. \ref{fig:model_performance} (left) illustrates the classification capability of the full model, with an AUC of 0.9444. This high score suggests that the model is highly effective at distinguishing between spike and non-spike periods. The curve’s steep ascent near the y-axis reflects high true positive rates even when the false positive rate is low, an important property for real-world use.

Fig. \ref{fig:model_performance} (right) shows the corresponding confusion matrix, confirming the model’s balanced performance. The model successfully identifies both classes, with only a few false positives and no false negatives which is encouraged when dealing with such problems where the stakeholders cannot afford a sudden spike while they can tolerate a false alarm. This is particularly encouraging, given the typical class imbalance in commodity datasets, where spike events are rare.

\begin{figure}[H]
    \centering
    \includegraphics[width=0.49\textwidth]{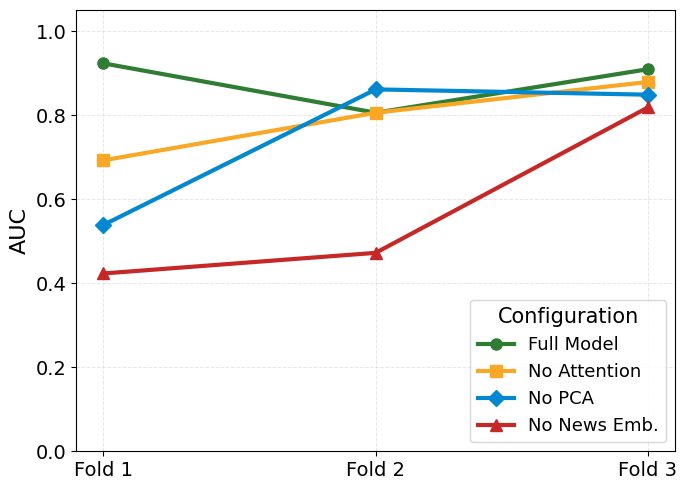}
    \hfill
    \includegraphics[width=0.49\textwidth]{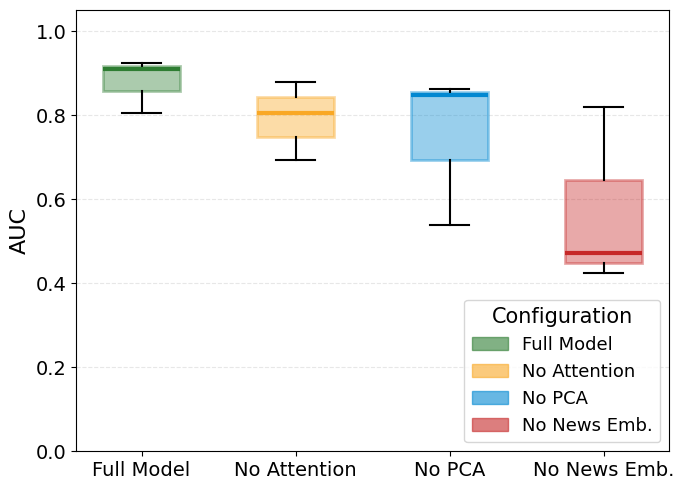}
    \caption{Ablation results. Left: Fold-wise AUC for each model variant. Right: Boxplot showing distribution of AUC scores.}
    \label{fig:ablation_analysis}
\end{figure}

To understand the relative contribution of each model component, ablation experiments were performed. The left plot of fig. \ref{fig:ablation_analysis} shows the AUC scores across cross-validation folds for different model variants. The full model consistently outperforms all ablated versions, indicating that each component contributes to the final performance. Removing any single component—whether the attention mechanism, PCA preprocessing, or news embeddings leads to measurable degradation in forecasting accuracy.

The right plot shows the variability of AUC scores across folds. The full model demonstrates the narrowest spread, indicating strong generalization and robustness to data splits. In contrast, the “No News Embeddings” variant shows a wide AUC spread, reflecting poor generalization and high variance. This instability highlights the critical role of textual news information in improving both predictive power and consistency.

\begin{table}[H]
\centering
\caption{Mean AUC across model variants with standard deviation.}
\begin{tabular}{lcc}
\toprule
\textbf{Model} & \textbf{AUC (Mean $\pm$ Std)} \\
\midrule
\textbf{Full Model}             & \textbf{0.9094 $\pm$ 0.0692} \\
Without Attention               & 0.8263 $\pm$ 0.0889 \\
Without PCA                     & 0.7584 $\pm$ 0.1302 \\
Without News Embeddings         & 0.4641 $\pm$ 0.2400 \\
\bottomrule
\end{tabular}
\label{tab:ablation_auc}
\end{table}

Table~\ref{tab:ablation_auc} quantifies the impact of each component. Removing the attention mechanism leads to a moderate drop in AUC, showing that attention helps the model focus on relevant time windows within the sequence. Excluding PCA has a larger impact, which suggests that dimensionality reduction plays a key role in filtering noise and improving model convergence. The most significant drop occurs when news embeddings are excluded; in this case, the model’s AUC falls to 0.4641, below random chance in some folds. This drastic decrease reinforces the central role of text-based features in capturing market sentiment and external shocks that pure numerical inputs fail to represent.

\begin{table}[H]
\centering
\caption{Weighted performance metrics. $\Delta$F1 indicates drop relative to the full model.}
\resizebox{\textwidth}{!}{
\begin{tabular}{lccccc}
\toprule
\textbf{Model} & \textbf{Accuracy} & \textbf{Precision (W)} & \textbf{Recall (W)} & \textbf{F1 Score ($\Delta$)} \\
\midrule
Full Model             & \textbf{0.91} & 0.83 & \textbf{0.91} & \textbf{0.87 (–)} \\
Without Attention      & 0.53 & \textbf{0.94} & 0.53 & 0.64 ($\downarrow$ 0.23) \\
Without PCA            & 0.40 & 0.85 & 0.40 & 0.40 ($\downarrow$ 0.47) \\
Without News Embeddings& 0.21 & 0.05 & 0.21 & 0.08 ($\downarrow$ 0.79) \\
\bottomrule
\end{tabular}
}
\label{tab:weighted_metrics}
\end{table}

Table~\ref{tab:weighted_metrics} provides a more detailed view of model behavior using additional performance metrics. The full model shows the best balance between precision and recall, reflected in its highest F1 score. Interestingly, while the model without attention achieves the highest precision (0.94), it suffers from poor recall (0.53), indicating a tendency to underpredict positive cases. In applications where missing true spikes is more costly than generating a few false alarms, recall becomes a more important metric, further justifying the inclusion of the attention layer.

The PCA layer contributes to model reliability by removing redundant or noisy features. Without PCA, the model’s recall and F1 scores drop sharply, and its overall accuracy becomes inconsistent. This suggests that careful preprocessing is crucial when dealing with high-dimensional embeddings, such as those derived from news data.

The most critical finding is the role of news embeddings. Removing this component causes the model to fail almost entirely, with a weighted F1 score of just 0.08 and accuracy of 0.21. This aligns with the idea that commodity markets are heavily influenced by external factors such as geopolitical news, weather reports, and economic policy which are often only available in unstructured text. Incorporating these signals allows the model to react more intelligently to context changes.

\begin{figure}[H]
	\centering
	\includegraphics[width=\textwidth]{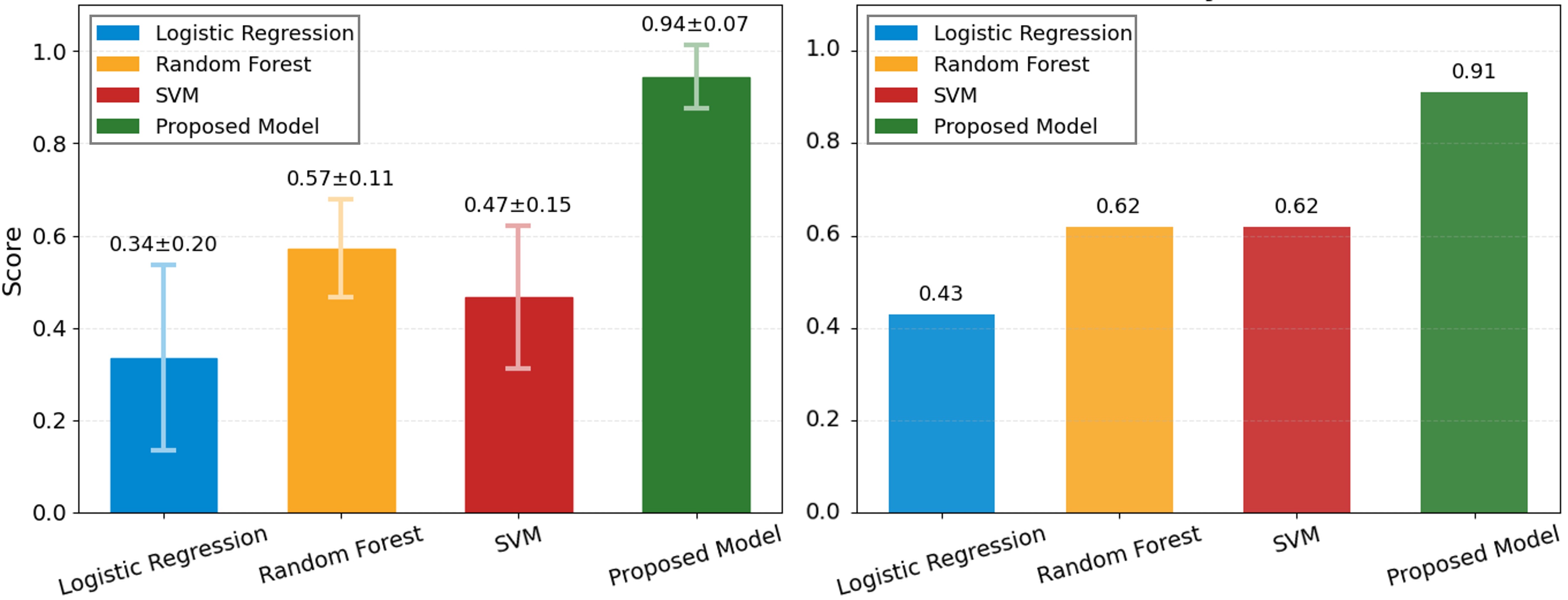}
	\caption{AUC(left) and Accuracy(right) comparison of baseline models vs. proposed approach}\label{modelscomp}
\end{figure}

The bar charts in fig. \ref{modelscomp} provide a comparative analysis of the proposed model against classical machine learning baselines including Logistic Regression, Random Forest, and SVM evaluated using a time-aware cross-validation scheme. Specifically, TimeSeriesSplit is employed to ensure that each fold respects the chronological order of the data, preventing information leakage and better reflecting real-world forecasting scenarios.

The left figure presents mean AUC scores across folds, where the proposed model significantly outperforms all baselines, achieving an average AUC of 0.94$\pm$0.07. In contrast, Random Forest, SVM, and Logistic Regression yield considerably lower AUCs of 0.57$\pm$0.11, 0.47$\pm$0.15, and 0.34$\pm$0.20, respectively. This confirms the proposed model's superior ability to discriminate between spike and non-spike periods across various time-based data splits.

The right figure shows accuracy scores from the same cross-validation setting. The proposed model achieves the highest accuracy at 0.91, compared to 0.62 for both Random Forest and SVM, and 0.43 for Logistic Regression.

\begin{figure}[H]
	\centering
	\includegraphics[width=\textwidth]{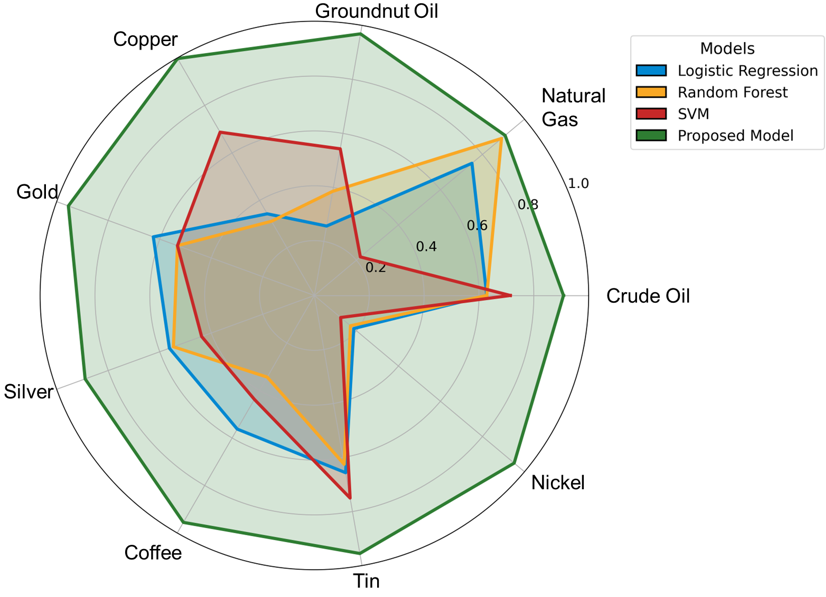}
	\caption{AUC comparison of forecasting models across a subset of individual commodities}\label{auc_comm}
\end{figure}
The individual commodity forecasting experiment is conducted to rigorously evaluate the proposed model’s capacity to generalize across heterogeneous market structures by measuring its predictive accuracy separately for each commodity, thereby validating its robustness in capturing commodity-specific temporal and semantic patterns.

The AUC results across individual commodities clearly demonstrate the robustness and generalizability of the proposed model. Unlike traditional classifiers such as Logistic Regression, Random Forest, and Support Vector Machines, which exhibit inconsistent performance across commodities ranging from high scores in certain cases (e.g., Natural Gas with Random Forest at 0.8906) to poor scores in others (e.g., Nickel with SVM at 0.1250) while the proposed model consistently achieves high AUC values for all commodities, with scores exceeding 0.88 across the board. This stability highlights the model’s ability to effectively capture underlying temporal and semantic patterns in the data, making it a reliable forecasting tool for diverse commodity markets. The close range of AUC values also suggests that the proposed framework maintains predictive strength regardless of the specific commodity being analyzed, which is critical for real-world deployment in heterogeneous economic contexts.

In summary, the results clearly demonstrate that each component of the proposed model contributes to its strong performance. The integration of time-series price trends, compressed news information, and attention-based temporal weighting enables robust and interpretable forecasts. 

\section{Conclusion}
\label{sec:conclusion}
This paper presents a novel multimodal forecasting framework that brings together structured time-series data and unstructured textual signals to improve the early detection of commodity price spikes. The core contribution lies in the integration of an agentic generative AI pipeline with a deep learning architecture that uses dual-stream LSTM networks and attention mechanisms. The generative AI agent autonomously retrieves, summarizes, and fact-checks relevant economic news from major global sources, producing concise yearly summaries that are semantically embedded and temporally aligned with commodity price data. These textual signals, when fused with historical price trends, allow the model to capture latent market drivers such as geopolitical conflicts, policy changes, and global disruptions that traditional models often overlook. The attention mechanism enables the model to focus on the most relevant patterns across both modalities, enhancing its ability to forecast rare but impactful spikes in commodity prices. The framework is rigorously evaluated over 64 years of data and demonstrates strong predictive performance, outperforming classical machine learning methods and ablated variants of the model. Collectively, this approach offers a more context-aware, adaptive forecasting tool that can support policy-making and risk management in volatile economic environments, particularly in countries where timely and accurate insights are critical for financial planning and stability.

Despite these promising results, the study has limitations. The analysis is based on yearly data, which limits the model’s responsiveness to short-term events. Furthermore, the relatively small number of positive spike examples increases the risk of overfitting, even with regularization and dimensionality reduction. Lastly, while the generative agent ensures factual accuracy through a verification process, the reliance on summary-level news may omit nuanced developments occurring within each year.

Future work will address these limitations by applying the framework to higher-frequency data, such as monthly or weekly commodity prices, and incorporating real-time news streams. Improvements in temporal alignment between price movements and textual inputs could enhance the model’s short-term predictive capabilities. Additionally, expanding the agentic pipeline to include multimodal inputs such as social media sentiment or policy announcements may further improve sensitivity to emerging market risks. Ultimately, the goal is to build an adaptable, real-time decision support system for commodity risk management in data scarce and economically vulnerable settings.

%% For numbered reference style

%\bibliographystyle{elsarticle-num} 
%\bibliography{references}


\end{document}